\documentclass[aps,twocolumn,showpacs]{revtex4}
\makeatletter

\newcommand{\Rmnum}[1]{\expandafter\@slowromancap\romannumeral #1@}
\makeatother
\usepackage{graphicx}
\usepackage{dcolumn}
\usepackage{bm}
\usepackage{CJK}
\usepackage{amsfonts}
\usepackage{psfrag}
\usepackage{wrapfig}
\usepackage{subfigure}
\usepackage{makeidx}
\usepackage{multirow}
\usepackage{epsf}
\usepackage{hyperref}
\usepackage{amsmath}
\usepackage{cases}

\begin{document}

\title{Symmetric and asymmetric optical multi-peak solitons on a continuous wave background in the femtosecond regime}
\author{Chong Liu$^1$}
\author{Zhan-Ying Yang$^1$}
\author{Li-Chen Zhao$^{1}$}
\author{Liang Duan$^{1}$}
\author{Guangye Yang$^{2}$}
\author{Wen-Li Yang$^3$}
\address{$^1$School of Physics, Northwest University, Xi'an
710069, China}
\address{$^2$Department of Physics, Shanxi Medical University, Taiyuan, Shanxi, 030001, China}
\address{$^3$Institute of Modern Physics, Northwest University, Xi'an
710069, China}
\begin{abstract}
We study symmetric and asymmetric optical multi-peak solitons \textit{on a continuous wave background} in the femtosecond regime of a single-mode fiber.
Key characteristics of such multi-peak solitons, as the formation mechanism, propagation stability, and shape-changing collisions, are revealed in detail.
Our results show that this multi-peak (symmetric or asymmetric) mode could be regarded as a single pulse formed by a nonlinear superposition of a periodic wave and a single-peak (W-shaped or antidark) soliton.
In particular, a phase diagram for different types of nonlinear excitations on a continuous wave background including breather, rogue wave, W-shaped soliton, antidark soliton, periodic wave, and multi-peak soliton is established based on the explicit link between exact nonlinear wave solution and modulation instability analysis.
Numerical simulations are performed to confirm the propagation stability of the multi-peak solitons with symmetric and asymmetric structures.
Further, we unveil a remarkable shape-changing feature of asymmetric multi-peak solitons.
It is interesting that these shape-changing interactions occur not only in the intraspecific collision (soliton mutual collision) but also in the interspecific interaction (soliton-breather interaction). Our results demonstrate that each multi-peak soliton exhibits the coexistence of shape change and conservation of the localized energy of light pulse against the continuous wave background.
\end{abstract}
\maketitle
\section{Introduction}
Nonlinear waves on a continuous wave background in optical fibers becomes a subject of intense research in both theory and experiment
these days \cite{n1,n2,n3,n4,n5,n6,n7,n8}. Especially, significant progress has been made on the
experimental verification of some unique nonlinear wave structures, including the Peregrine rogue waves \cite{rw}, Kuznetsov-Ma breathers (KMBs) \cite{km}, Akhmediev breathers (ABs) \cite{ab}, as well as their shape-unchanging interactions such as
the AB collision \cite{abc}, and superregular breather \cite{n8,sr} (i.e., quasi-AB collision with $\pi/2$ phase shift) in the picosecond regime.
These picosecond pulses are well described by the standard nonlinear Schr\"{o}dinger equation (NLSE) which
accounts for the second-order dispersion and self-phase modulation.
Specifically, these unique waves appear as a result of the modulation instability (MI) \cite{n5,n6,n7,n8,MI1}, and in turn the common features and differences among
wave manifestations enrich the MI understanding of the nonlinear stage. It should be emphasized that a crucial precise link between rogue waves and zero-frequency MI subregion is recently unveiled \cite{MI2,MI3}, although a comprehensive investigation of exact relations between various types of nonlinear waves on a background and MI still remains largely unexplored.
On the other hand, the utility of these waves based on their special properties in generating high-quality pulse trains \cite{u1}, high-power pulses \cite{u2},
breatherlike solitons \cite{u3}, nonlinear Talbot effects \cite{u4}, and the Peregrine comb \cite{u5} has been revealed.

In the present work we extend nonlinear waves on a continuous wave background in the femtosecond regime, since ultrashort pulses
are tempting and desirable to improve the capacity of high-bit-rate transmission systems. However in this case, higher-order effects such as the higher-order dispersion and self-steepening play an important role and become non-negligible \cite{h1,h2,h3}. The resulting models of higher-order NLSE thus describe nonlinear waves on a continuous wave background with higher accuracy than the standard NLSE. Recent studies demonstrated that nonlinear waves on a continuous wave background in the femtosecond regime exhibit structural diversity beyond the reach of the standard NLSE \cite{f1,f2,f3,f4,f5,f6,f7,f8,f9,f10,f11,f12,f13,f14,f15}; in particular, some interesting types of nonlinear waves on a background which completely differ from the known breathers and rogue waves have been unveiled \cite{f7,f8,f9,f10,f11,f12,f13,f14,f15}.
The underlying mechanism can be qualitatively but quite explicitly interpreted by the corresponding MI features \cite{f7,f10,f11}.
Here, we report and discuss a family of multiparametric symmetric and asymmetric multi-peak solitons on a continuous wave background in the femtosecond regime.
One key characteristic of this soliton is that it exhibits both localization and periodicity along the transverse distribution on a background; the corresponding periodicity and localization for (symmetric or asymmetric) solitons are well described by a periodic wave and a single-peak (W-shaped or antidark) soliton, respectively.
Especially, we focus our attention on the important properties of this multi-peak solitons, including the generation mechanism, propagation stability, and the shape-changing interaction feature of asymmetric multi-peak solitons, which have not been studied so far, to our knowledge.
For special parameter values (the continuous wave background frequency vanishes), a part of our general multi-peak soliton solution (i.e., the symmetric case) reduces to results reported quite recently \cite{f13}.

The rest of the paper is structured as follows. In Section \uppercase \expandafter {\romannumeral 2}, the exact solution with a unified form describing symmetric and asymmetric multi-peak solitons on a background of the higher-order NLSE is obtained. The consistency between the symmetric and asymmetric solitons is revealed by the optical intensity against the background. Section \uppercase \expandafter {\romannumeral 3} shows that the periodicity and localization for (symmetric or asymmetric) solitons are well described by a periodic wave and a single-peak (W-shaped or antidark) soliton. In Section \uppercase \expandafter {\romannumeral 4}, the phase diagram for different types of nonlinear excitations is presented on the continuous wave frequency and perturbed frequency plane. Numerical simulations were performed timely to confirm the propagation stability of the multi-peak solitons in Section \uppercase \expandafter {\romannumeral 5}. Section \uppercase \expandafter {\romannumeral 6} unveils the striking shape-changing feature of asymmetric multi-peak solitons which occur not only in the intraspecific collision (soliton mutual collision) but also in the interspecific interaction (soliton-breather interaction). The final section presents our conclusions.

\section{The model and multi-peak solitons on continuous wave background}
Femtosecond pulses (i.e., the duration is shorter than 100 fs) propagation in optical fibers with higher-order physical effects such as the third-order dispersion,
self-steepening, and delayed nonlinear response is governed by the following higher-order NLSE \cite{h1,h2}
\begin{eqnarray}
\label{equ1}
iu_\xi+\frac{1}{2}\alpha u_{\tau\tau}+\gamma|u|^2u&-&i\beta u_{\tau\tau\tau}-is(|u|^2u)_\tau\nonumber\\
&-&i\delta u(|u|^2)_\tau=0,
\end{eqnarray}
where $u$ is the envelope of the electric field, $\xi$ is the propagation distance, $\tau$ is the retarded time,
$\alpha$ and $\gamma$ are second-order dispersion and self-phase modulation,
$\beta$ is the third-order dispersion, $s$ is the self-steepening coefficient, and $\delta$
is the delayed nonlinear response. All quantities have been normalized.

To study nonlinear waves in the femtosecond regime exactly,
we shall consider a special parametric condition for the higher-order terms, i.e., $s=6\beta$, $s+\delta=0$ with $\alpha=\gamma$.
As a result, the model (\ref{equ1}) reduces to the integrable Hirota equation \cite{Hirota}.
The latter has been studied in a number of papers \cite{f3,f10,f13,Hirota,s1,s2,s3,s4,b}, which involve mainly the standard bright solitons \cite{s1,s2,s3,s4}, breathers \cite{b}, and rogue waves \cite{f3}.
In contrast to the aforementioned results, we introduce, in the following, an interesting family of multiparametric nonlinear wave solutions describing symmetric and asymmetric multi-peak solitons on a continuous wave background in the femtosecond regime.

By means of the Darboux transformation method \cite{DT1,DT2}, multi-peak solitons on a continuous wave background with symmetric and asymmetric amplitude structures can be described by the analytical unified solution with a general and concise form,
\begin{equation}
\label{equ2}
u_{1,2}=\left[\frac{\Delta_{1,2}\cosh(\varphi+\delta_{1,2})+\Xi_{1,2}\cos(\phi+\xi_{1,2})}
{\Omega_{1,2}\cosh(\varphi+\omega_{1,2})+\Gamma_{1,2}\cos(\phi+\gamma_{1,2})}+a\right]e^{i\theta},
\end{equation}
where the continuous wave background has the expression: $u_0=ae^{i\theta}$, $\theta=q \tau+[\alpha a^2-\alpha q^2/2+\beta(6qa^2-q^3)]\xi$,
and $u_1$, $u_2$ stand for the symmetric and asymmetric multi-peak solitons, respectively.
It is evident that the solution (\ref{equ2}) is formed by a nonlinear superposition of the hyperbolic function $\cosh{\varphi}$ and trigonometric function $\cos{\phi}$ on the background $u_0$.
This unique nonlinear superposition signal exhibits the characteristics of the nonlinear structures on the nonvanishing background, which is expressed as:
\begin{eqnarray}
\label{equ3}
&&\varphi=2\eta_i(\tau + v \xi),~~~~~~\phi = 2\eta_r(\tau + v \xi),\nonumber\\
&&v=\beta(2 a^2+4 a_1^2- q_1^2)-(q_1+q)(q\beta+\alpha/2),\nonumber\\
&&\eta_r+i\eta_i=\sqrt{\epsilon+i\epsilon'},~~\epsilon=a^2-a_1^2 + (q-q_1)^2/4,\nonumber\\
&&\epsilon'=a_1(q-q_1),~~q_1=-\alpha/(4\beta)-q/2,
\end{eqnarray}
with the corresponding amplitude and phase notations:
\begin{eqnarray}
\Delta_{1}&=&-4aa_1\sqrt{\rho+\rho'},~~~\Delta_{2}=-4a^2a_1,\nonumber\\
\Xi_{1}&=&2a_1\sqrt{\chi^2-(2a^2-\chi)^2},~\Xi_{2}=4aa_1\sqrt{2(i\epsilon'-\epsilon)},\nonumber\\
\Omega_{1}&=&\rho+\rho',~~~~~\Omega_{2}=\sqrt{\rho^2-\rho'^2},\nonumber\\
\Gamma_{1}&=&-2a(\eta_i+a_1),~~~~~\Gamma_{2}=\sqrt{\varrho^2+\varrho'^2},\nonumber\\
\delta_1&=&\textrm{arctanh}(-i\chi_1/\chi_2),\nonumber\\
\delta_2&=&\textrm{arctanh}[i2(\eta_i+i\eta_r)/(q-q_1-2ia_1)],\nonumber\\
\xi_1&=&-\textrm{arctan}[i(2a^2-\chi)/\chi],\nonumber\\
\xi_2&=&-\textrm{arctan}[i2(\eta_i+i\eta_r)/(q-q_1-2ia_1)],\nonumber\\
\omega_1&=&0,~~~~~\omega_2=\textrm{arctanh}(-\rho'/\rho),\nonumber\\
\gamma_1&=&0,~~~~~\gamma_2=-\textrm{arctan}(\varrho'/\varrho),\nonumber
\end{eqnarray}
and $\rho=\epsilon+ 2a_1^2 + \eta_i^2 + \eta_r^2$, $\rho'=\eta_r(q-q_1)+2\eta_i a_1$, $\varrho=\epsilon+ 2a_1^2 - \eta_i^2 - \eta_r^2$, $\varrho'=\eta_i(q_1- q)+2\eta_r a_1$, $\chi=\chi_1^2 + \chi_2^2 + a^2$, $\chi_1=\eta_r+(q-q_1)/2$, $\chi_2 =\eta_i + a_1$.

\begin{figure}[htb]
\centering
\subfigure{\includegraphics[height=42mm,width=42mm]{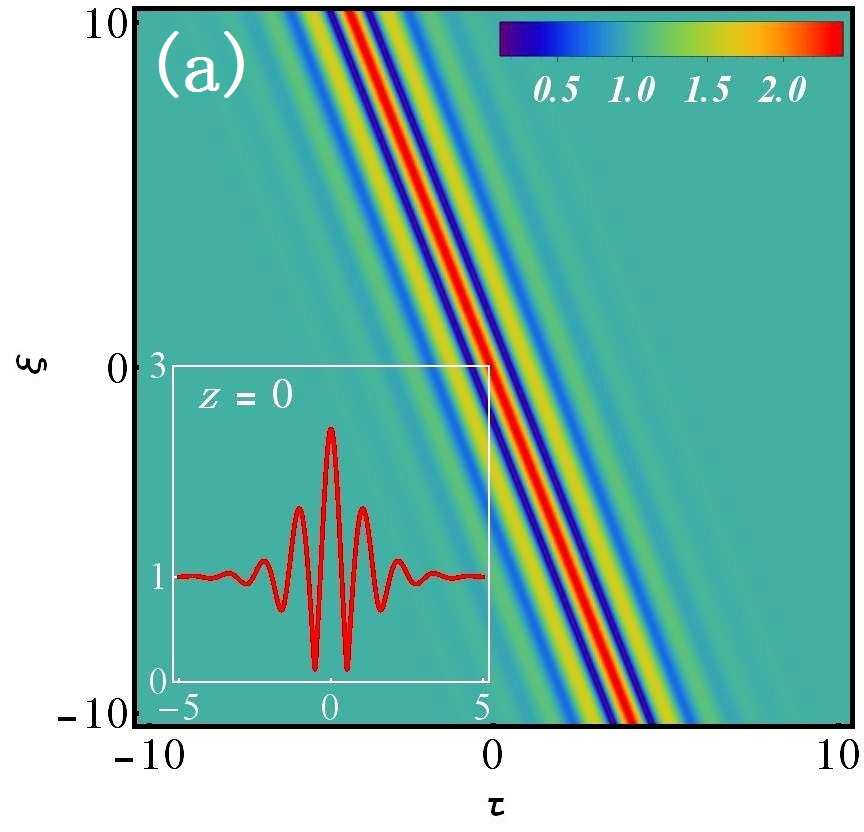}}
\hfil
\subfigure{\includegraphics[height=42mm,width=42mm]{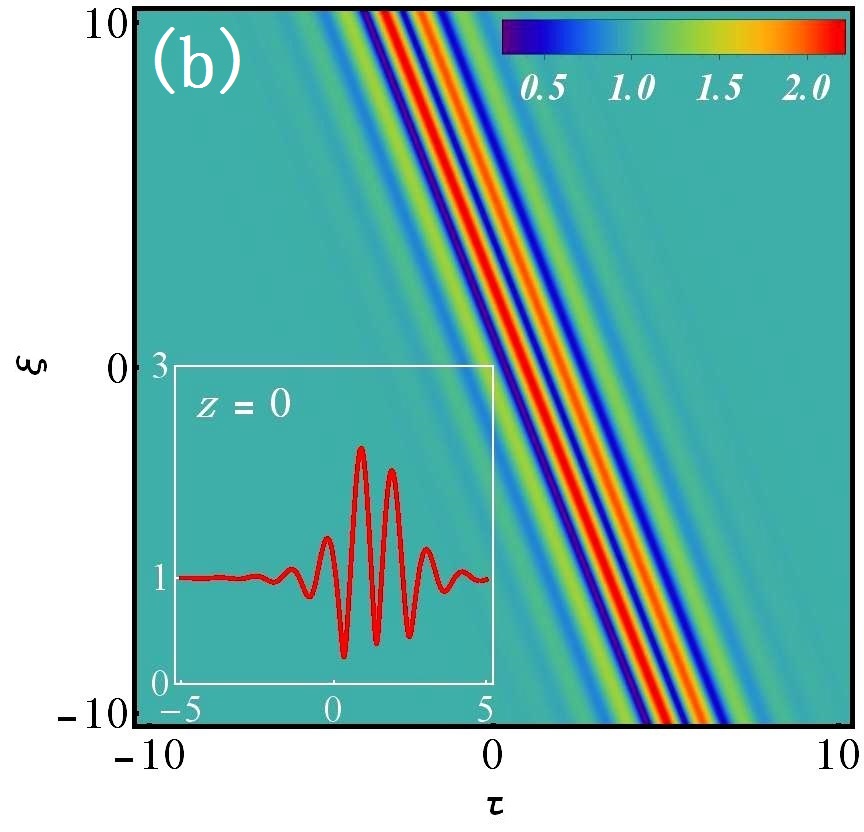}}
\caption{(color online) Optical amplitude distributions $|u|$ of multi-peak solitons on a continuous wave background with (a) symmetric $u_1$ and (b) asymmetric $u_2$ structures.
The setup is $q=3q_s$, $a=1$, $a_1=0.7$, $\beta=0.1$, $\alpha=1$, with $q_s=-\alpha/(6\beta)$.}\label{fig1}
\end{figure}

\begin{figure}[htb]
\centering
\subfigure{\includegraphics[height=42mm,width=42mm]{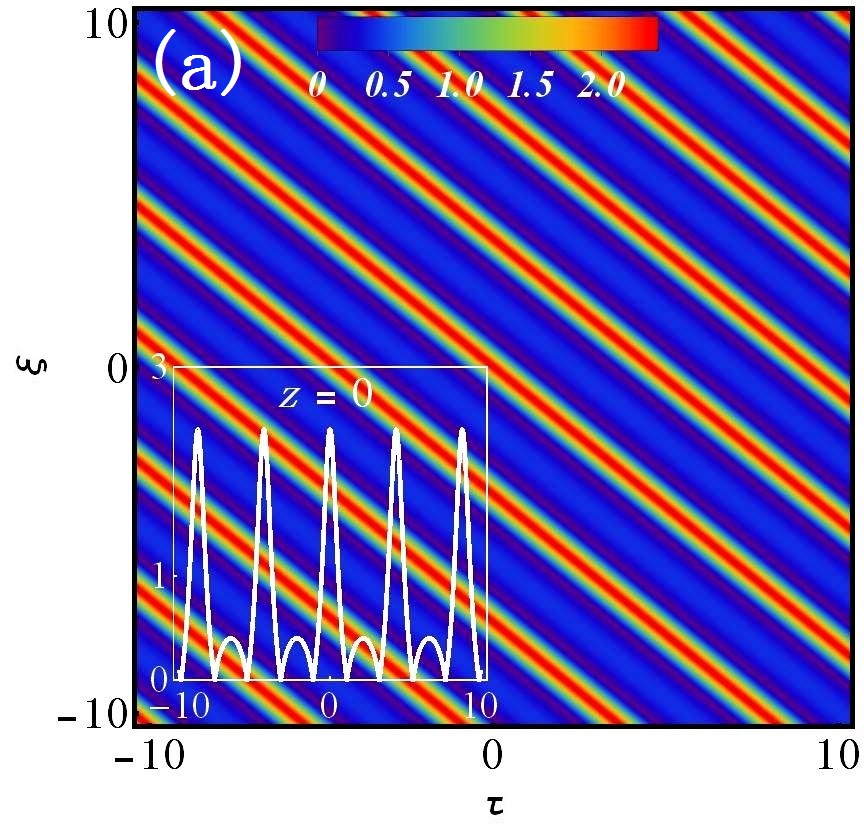}}
\hfil
\subfigure{\includegraphics[height=42mm,width=42mm]{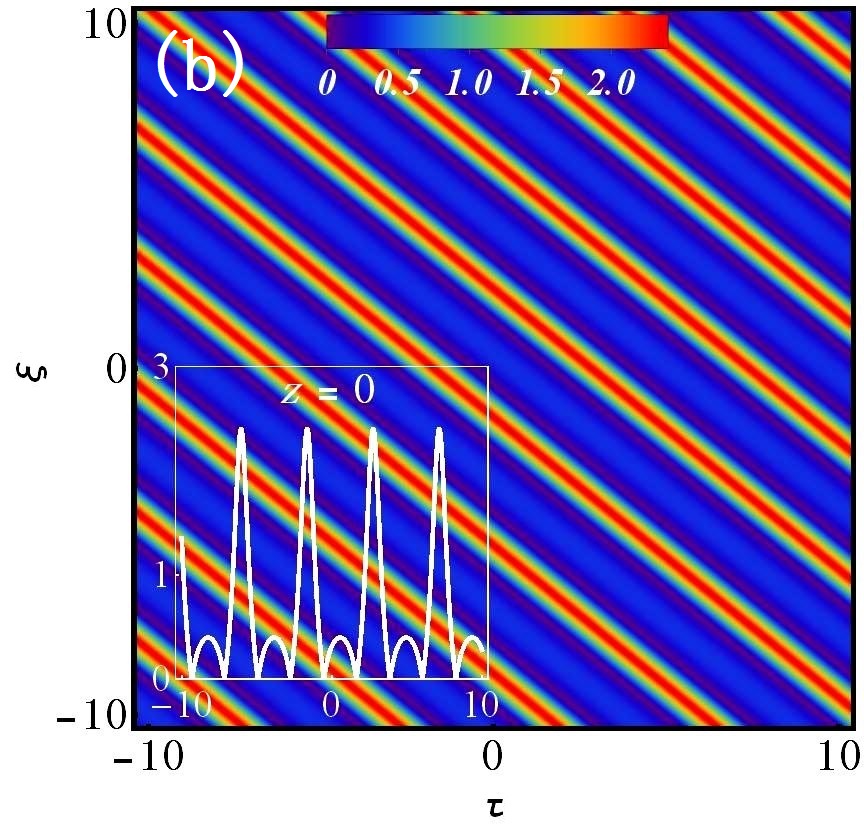}}
\hfil
\subfigure{\includegraphics[height=42mm,width=42mm]{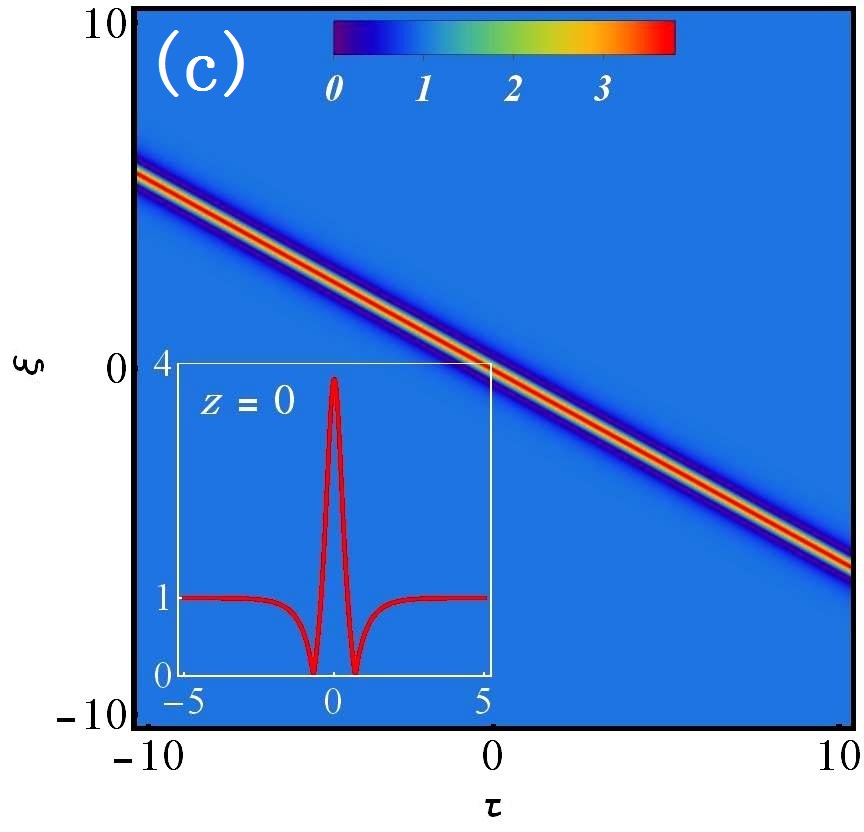}}
\hfil
\subfigure{\includegraphics[height=42mm,width=42mm]{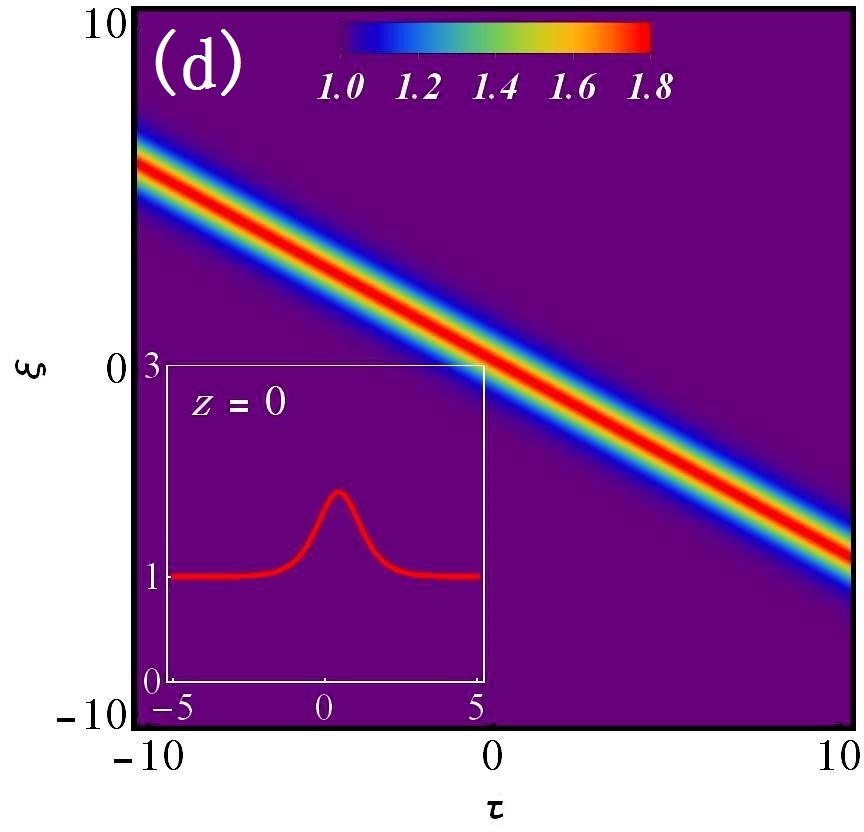}}
\caption{(color online) Periodic waves and single-peak solitons extracting from the multi-peak solitons.
Periodic waves: (a) $u_{p1}$ and (b) $u_{p2}$ with $a_1=0.7$, given by Eqs. (\ref{equ4});
single-peak solitons: (c) $u_{s1}$ and (b) $u_{s2}$ with $a_1=1.4$, given by Eqs. (\ref{equ5}).
Other parameters are $a=1$, $\beta=0.1$, $\alpha=1$.}\label{fig2}
\end{figure}

The above expressions depend on the continuous wave background amplitude $a$, frequency $q$, the real constant $a_1$ (without loss of generality we let $a_1\geq0$),
the real parameter $\alpha$ describing the group-velocity dispersion and the self-phase modulation,
the real parameter $\beta$ (a small value) which is responsible for the higher-order terms.
Note that the existence condition $q_1=-\alpha/(4\beta)-q/2$ [see Eq. (\ref{equ3})] implies that this nonlinear mode is induced by the higher-order effects ($\beta\neq0$) and, therefore, it has no analogy in the picosecond regime governed by the standard NLSE.
Additionally, the background frequency $q$ plays a key role in properties of nonlinear modes, since it cannot be ignored by the Galilean transformation. Indeed, the nonlinear modes exhibit prolific structures depending on the value of $q$ [f.i., see Figs. \ref{fig1} and \ref{fig2}].

The key novel property of this solution is that it features both hyperbolic and trigonometric with the same velocity $v$ [see the hyperbolic function ($\cosh{\varphi}$) and trigonometric function ($\cos{\phi}$)].
This implies $\eta_r\neq0$, $\eta_i\neq0$ [thus $q\neq q_1$, i.e., $q\neq-\alpha/(6\beta)$]. As a result, the hyperbolic functions and trigonometric functions describe the localization and the periodicity of the transverse distribution $\tau$ of the localized waves, respectively.
The corresponding typical optical amplitude ($I=\sqrt{|u_{1,2}(\xi,\tau)|^2}$) profiles are well displayed in Fig. \ref{fig1}.

Remarkably, the pulse features a localized soliton-like multi-peak structure propagating in $\xi$ direction. Or, strictly speaking, this interesting solitary mode exhibits both localization and periodicity along the transverse distribution $\tau$. This indicates that the characteristics of this wave come from a mixture of a soliton and a periodic wave.
Specifically, this pulse possesses a main peak and several sub-peaks, and the latter are distributed on both side of the main peak in a symmetric or asymmetric way. Although the maximum optical intensity is different,
the interesting connection is that, the optical intensity against the background of the two solitons with symmetric and asymmetric profiles, turns out to coincide with each other, i.e.,
$\int_{-\infty}^{+\infty}\left(|u_{1}|^2-a^2\right)\textrm{d}\tau=\int_{-\infty}^{+\infty}\left(|u_{2}|^2-a^2\right)\textrm{d}\tau$.
It should be noted that the solution (\ref{equ2}) includes, as a special case, the symmetric solution $u_1$ with $q=0$ that was reported in \cite{f13}.

\section{Formation mechanism for multi-peak solitons}
Next, to further understand the formation mechanism of the localized multi-peak structure on a background, our attention is then focused on extracting separately the periodic wave (periodicity) and soliton (localization) from the mixed state.

To this end, the nonlinear excitation signal possesses a single modulated function (hyperbolic or trigonometric functions). In this interesting case, the background frequency $q$ plays a key role and should be chosen as $q=-\alpha/(6\beta)$ (thus $q\neq0$). Specifically, the periodic wave exists on its own when $\eta_i=0$ vanishes (thus $a_1<a$), while the soliton exists in isolation when $\eta_r=0$ vanishes (thus $a_1>a$). The unified explicit expressions read, for the periodic wave,
\begin{equation}
\label{equ4}
u_{p~1,2}=\left[\frac{2\eta^2}{a-e^{i\sigma}a_1\cos[2\eta(\tau+v\xi)-\mu]}-a\right]e^{i\theta},
\end{equation}
where $\eta=\pm\sqrt{a^2-a_1^2}$, $\sigma=\sigma_{1,2}=\{0,\pi\}$ with $\mu=\mu_{1,2}=\{0,\arctan(-\eta_r/a_1)\}$, and for the soliton
\begin{equation}
\label{equ5}
u_{s~1,2}=\left[\frac{2\eta'^2}{e^{i\sigma}a_1\cosh[2\eta'(\tau+v\xi)+\mu']-a}-a\right]e^{i\theta},
\end{equation}
where $\eta'=\pm\sqrt{a_1^2-a^2}$, $\sigma=\sigma_{1,2}=\{0,\pi\}$ with $\mu'=\mu'_{1,2}=\{0,\textrm{arctanh}(-\eta_i/a_1)\}$.

Figure \ref{fig2} depicts the typical optical amplitude profiles of the periodic waves and solitons extracting from the localized modes $u_1$, $u_2$, respectively.
As is shown, the periodic waves possess the same profile feature (the same W-shaped structure of periodic units and the same intensity), but for a slight phase shift.
Surprisingly, the single-peak solitions display completely different structures depending on the phases: one is a W-shaped soliton with one peak and two symmetric valleys whose central position is located at $(\xi,\tau)=(0,0)$, extracting from the symmetric multi-peak soliton $u_1$; the another is an antidark soliton (a bright soliton on a nonvanishing background) with a slight phase shift, extracting from the asymmetric multi-peak soliton $u_2$. We note that the optical intensity against the background of the two types of solitons is consistent with each other. It is worth pointing out that, the rational W-shaped soliton reported before, is only the limiting case of $u_{s1}$ and $u_{p1}$ with $a_1=a$ \cite{f10}.

An analysis of above results, shown in Figs. \ref{fig1} and \ref{fig2}, implies that, a formation mechanism of the localized periodic modes may be interpreted as a nonlinear superposition of a periodic wave and a single-peak soliton. The W-shaped and antidark solitons with periodic modulation give rise to the symmetric and asymmetric multi-peak modes, respectively.

\section{phase diagram of various nonlinear excitations}

\begin{figure}[htb]
\centering
\subfigure{\includegraphics[height=42mm,width=42mm]{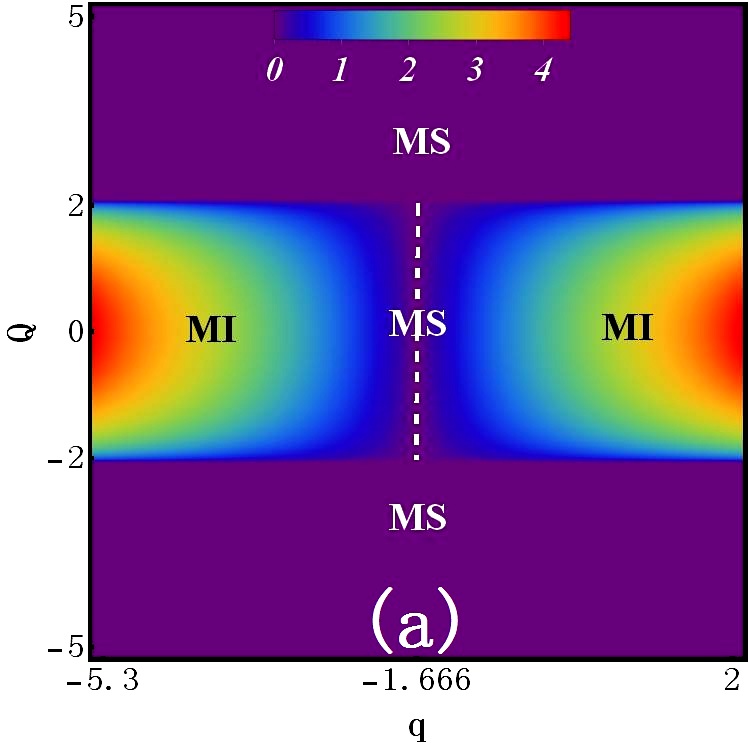}}
\hfil
\subfigure{\includegraphics[height=42mm,width=42mm]{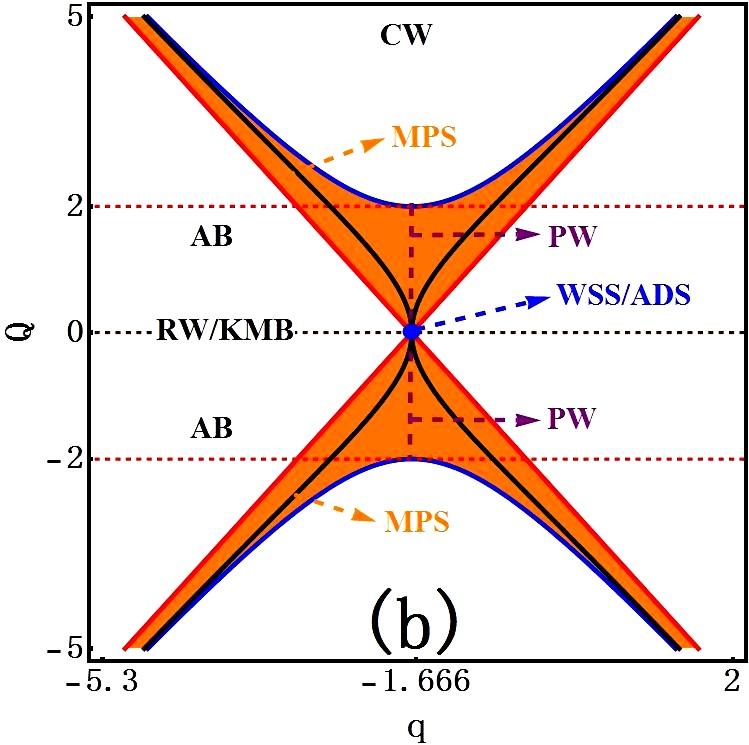}}
\caption{(color online) (a) Characteristics of modulation instability growth rate on perturbed frequency and background frequency ($Q,q$) plane with higher-order effects, and (b) the explicit correspondence of phase diagram for different types of nonlinear excitations on a continuous wave background. In (a), notations ``MI'' and ``MS'' denote modulation instability and modulation stability regions, respectively. In (b), ``RW'' (Peregrine rogue wave), ``KMB'' (Kuznetsov-Ma breather), and ``AB'' (Akhmediev breather) are casted in the MI region with corresponding exact perturbed frequencies $Q=0$, $Q=0$, and $|Q|<2a$ ($Q\neq0$), respectively. ``WSS'' (W-shaped soliton), ``ADS'' (antidark soliton), and ``PW'' (periodic wave) are mapped in the MS line $q=q_s$; specifically, W-shaped and antidark solitons exist at $(q,Q)=(q_s,0)$ (the blue dot), and periodic waves exist at $q=q_s$ with $|Q|\leq2a$ ($Q\neq0$). The ``MPS'' (multi-peak soliton), whether symmetric or asymmetric structure, is displayed in
the same orange ``X-shaped'' region, which is depicted via the explicit perturbed frequency expression $2\eta_r$ with different values of $a_1$, from $a_1=0$ to $a_1\gg a$. Here the solid blue, black, red lines in the orange region correspond to the cases with $a_1=0$, $a_1=a$, $a_1\gg a$, respectively.
The setup is $\beta=0.1$, $\alpha=1$, $a=1$.}\label{fig3}
\end{figure}

The next step of interest and significance is to understand the multi-peak structures on a background via MI.
To this end, we first turn our attention to the standard linear stability analysis of the continuous wave $u_0$ via adding small amplitude perturbed Fourier modes $p$ i.e., $u_{p}=[a+p]e^{i\theta}$, where
$p=f_+e^{i(Q\tau+\omega \xi)}+f_{-}^{*}e^{-i(Q\tau+\omega^* \xi)}$, $f_+$, $f_{-}^{*}$ are small amplitudes, $Q$ represents perturbed frequency, and
the parameter $\omega$ are assumed to be complex. In this case, by comparing the perturbed modes $p$ with the unique exact nonlinear superposition signal in $u_{1,2}$,
the nonlinear modes, Eq. (\ref{equ2}), shall be regarded as the nonlinear perturbation signal on background $u_0$ with the explicit perturbed frequency $Q=2\eta_r$.
Different perturbed frequencies correspond to different types of nonlinear excitations \cite{MI4}.
In this regard, we can establish an explicit correspondence between various nonlinear excitations and MI on the ($Q,q$) plane.
The corresponding characteristic outcomes are displayed in Fig. \ref{fig3}.

Figure \ref{fig3}(a) depicts the typical characteristics of MI growth rate $G=-\textrm{Im}\{\omega\}$, and Fig. \ref{fig3}(b) shows the explicit correspondence between MI and various nonlinear excitations. Remarkably, there is a modulation stability (MS) region [dashed line in Fig. \ref{fig3}(a)], i.e., $G=0$, in low perturbed frequency region ($|Q|\leq 2a$), resulting from the higher-order effects, which is given analytically $q=q_s=-\alpha/(6\beta)$. Hence MI is always present in the region $|Q|\leq2a$ ($Q\neq0$, $q\neq q_s$). Instead, the MS regions contain the whole higher perturbation frequency region ($|Q|>2a$) and a special MS region $q=q_s$ in a low perturbation frequency region ($|Q|\leq2a$).

An interesting finding is that the MS condition $q=q_s$, is consistent with the existence condition for the periodic wave and W-shaped/antidark soliton, Eqs. (\ref{equ4}) and (\ref{equ5}). According to their perturbed frequencies, the W-shaped/antidark solitons are located at $(q,Q)=(q_s,0)$, and periodic waves exist at $q=q_s$ line with $|Q|<2a$, $Q\neq0$.

Next, we map the whole distribution region of multi-peak solitons on the MI plane.
To this end, the modulated parameter $a_1$ of the perturbed frequency $2\eta_r$ is considered as a continuous variable from $a_1=0$ to $a_1\gg a$. The corresponding marginal condition, can be found analytically: $Q=\pm\{4a^2+9\Lambda^2/4\}^{1/2}$ (when $a'=0$ with $\Lambda=q-q_s$), $Q=\pm3|\Lambda|/2$ (when $a_1\gg a$). Thus,
the existence range of the multi-peak solitons is identified in the orange ``X-shaped'' region between the two marginal lines [see Fig. \ref{fig3}(b)].

\begin{figure*}[htb]
\centering
\includegraphics[height=65mm,width=110mm]{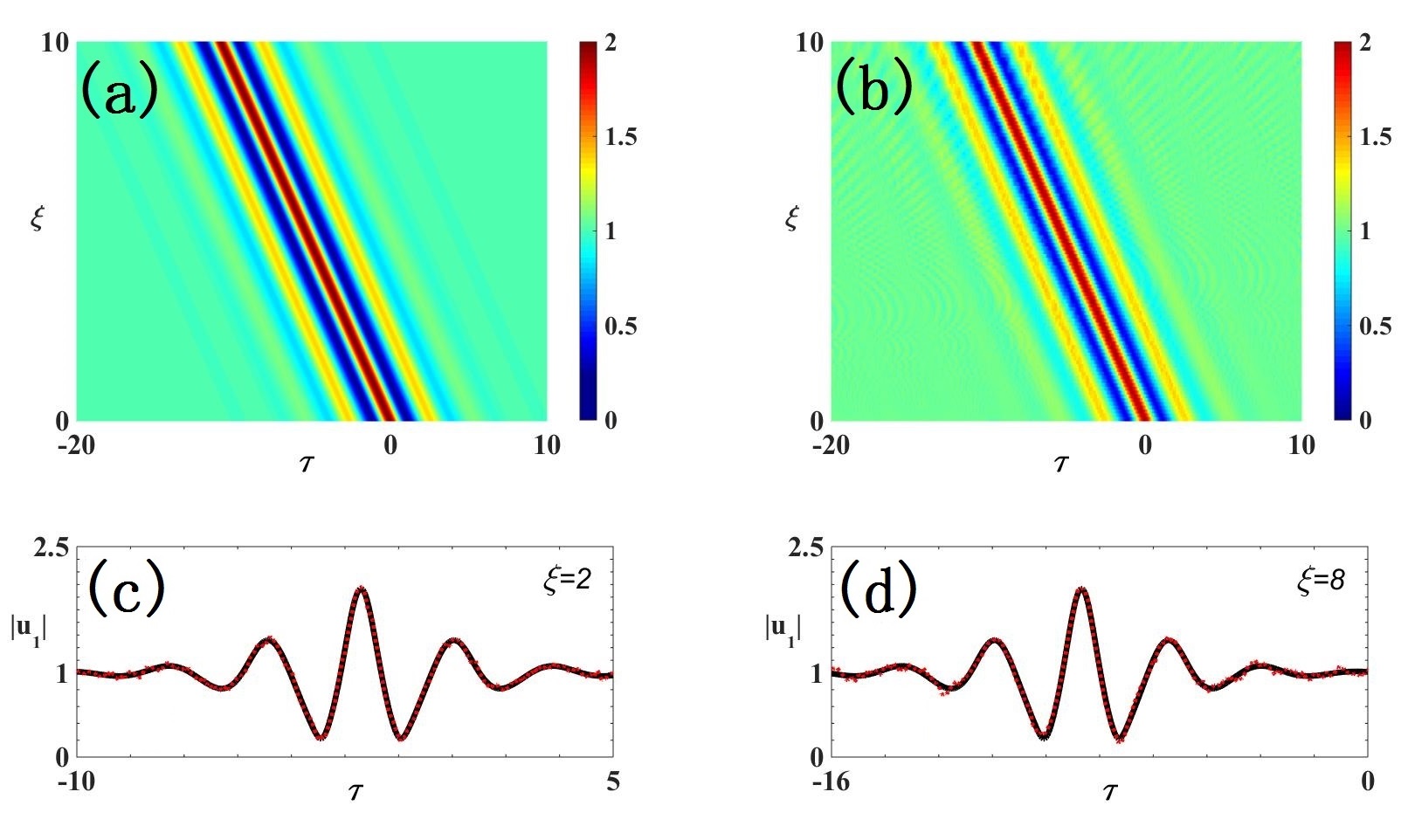}
\caption{(color online) Stability confirmation for the symmetric multi-peak solitons $|u_1(\xi,\tau)|$,  (a) the analytical solution, (b) the numerical simulations,
(c) and (d) the corresponding amplitude profiles of the analytical solution (solid black line) and the numerical simulation (dotted red line) at $\xi=2,8$. The setup is $q=3q_s/2$, $a=1$, $\alpha=1$, $\beta=0.1$, and $a_1=0.5$.}\label{fig4}
\end{figure*}
\begin{figure*}[htb]
\centering
\includegraphics[height=65mm,width=110mm]{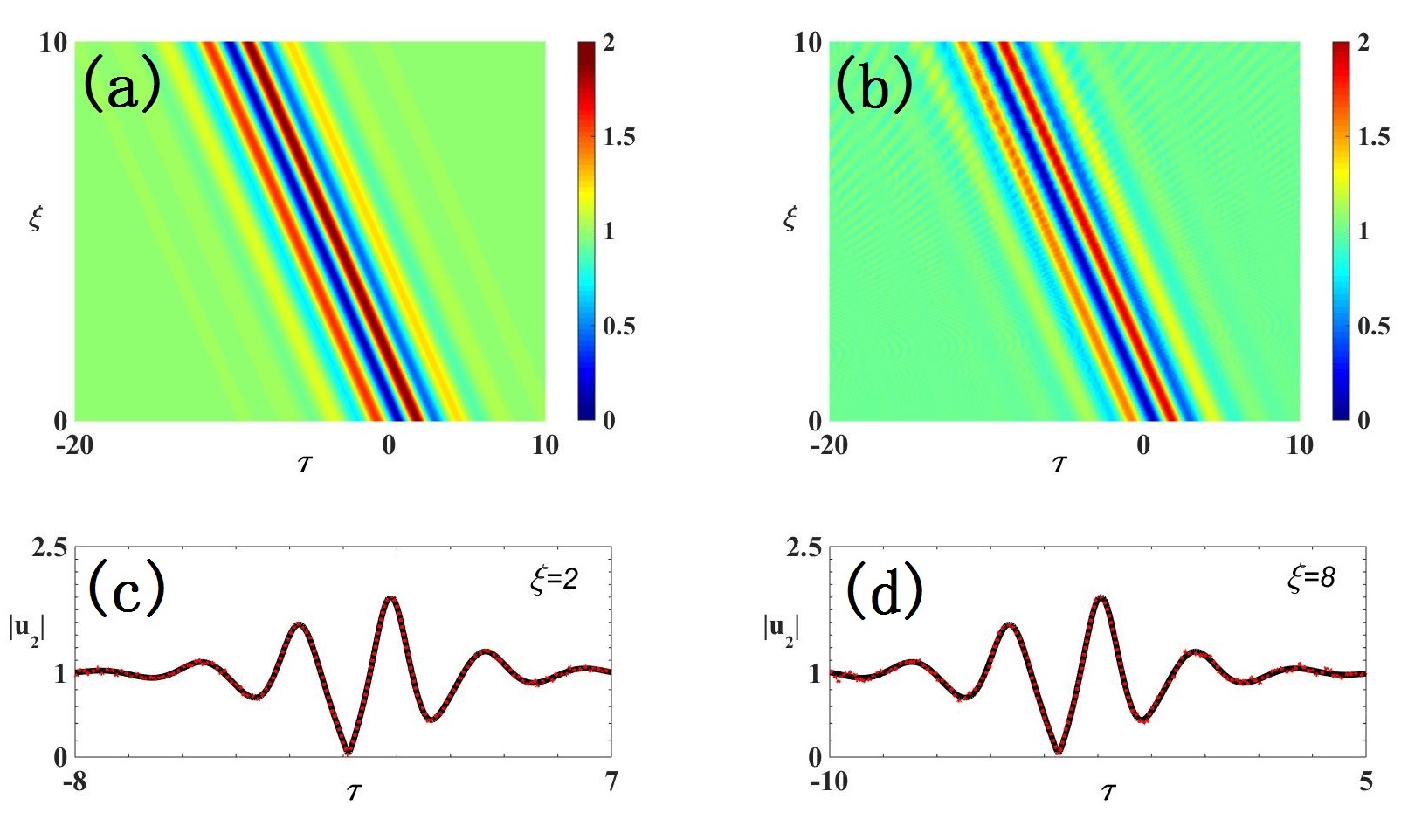}
\caption{(color online) Stability confirmation for the asymmetric multi-peak solitons $|u_2(\xi,\tau)|$. The setup is the same as in Fig. \ref{fig4}, but the initial state $u_{2}(0,\tau)$.}\label{fig5}
\end{figure*}

Remarkably, this unique ``X-shaped'' region involves both a MS subregion with higher perturbation frequency ($|Q|>2a$) and a MI subregion with low perturbation frequency ($|Q|<2a$). Further, the solid black line in the ``X-shaped'' region represents the intermediate condition with $a_1=a$, which can be given analytically as: $Q=\pm \{[(9\Lambda^2/8)^2+9a^2\Lambda^2]^{1/2}+9\Lambda^2/8\}^{1/2}$. It is interesting to note that, if $q=q_s$, the region $a_1\geq a$ [i.e., the range from red line to black line] reduces to the point $(q,Q)=(q_s,0)$, where the multi-peak solitons are converted into the W-shaped/antidark solitons; while another region $0\leq a_1<a$ [i.e., the range from black line to blue line] reduces to the $q=q_s$ line with $|Q|\leq2a$, $Q\neq0$, where the multi-peak solitons are translated into the periodic waves. We stress that this finding coincides with the transition condition for nonlinear waves in Eqs.(\ref{equ4}) and (\ref{equ5}).

\section{Stability for multi-peak solitons}

As is well known, the stability plays an irreplaceable role in nonlinear wave realization and application in experiment. On the other hand,
one should keep in mind that the multi-peak solitons reported above are on a continuous wave background. The latter, in general, displays the feature of MI, namely, a small perturbation may distort the wave profiles formed on top of it. In this regard, we shall test the stability of multi-peak soliton propagation on a continuous wave background with symmetric and asymmetric structures. We perform direct numerical simulations of Eq. (\ref{equ1}) by the split-step Fourier method with the initial condition $u_{1,2}(0,\tau)$, i.e., the exact solution (\ref{equ2}) at $\xi=0$.
Figures \ref{fig4} and \ref{fig5} show the stability of the symmetric and asymmetric multi-peak solitons by
comparison the analytical solution (\ref{equ2}) $u_{1,2}$ with the numerical simulation of Eq. (\ref{equ1}).

In Fig. \ref{fig4}, we first show the propagation stability of a typical symmetric multi-peak soliton $u_{1}$.
Interestingly, the numerical result [Fig. \ref{fig4} (b)] shows clearly that the symmetric multi-peak soliton can propagate in a stable manner against
the MI and ineradicable numerical deviations.
In particular, we compare the numerical result with the analytical solution by the
corresponding amplitude profiles at different $\xi$. It shows in Figs. \ref{fig4} (c) and (d) that the numerical result (dotted red line) is in good agreement with the
analytical solution (solid black line) at $\xi=2,8$.

Next, we test the stability of the asymmetric multi-peak soliton.
Here, we keep the the same parameters as in Fig. \ref{fig4}, but the initial state is chosen as $u_{2}(0,\tau)$.
Our results indicate no collapse arising from the MI and numerical deviations. Instead, stable propagation over tens
of propagation distances is observed (see Fig. \ref{fig5}). Namely, the asymmetric feature of the transverse amplitude distribution seems to have no effect on the propagation stability of the multi-peak solitons.
It should be pointed out that although here we have demonstrated the
results of the stability only for the multi-peak solitons in Eq. (\ref{equ1}), similar
conclusions hold for other-type nonlinear waves as well, provided
propagation distances are kept within reasonable values.

\section{shape-changing feature of asymmetric multi-peak solitons}

Let us then, in this section, focus our attention on characteristics of interaction between nonlinear waves reported above.
In fact, we find that rich intraspecific and interspecific interactions do exist in this system via extracting
explicit existence conditions from Table \uppercase \expandafter {\romannumeral 1} in the Appendix. However,
our interest is confined to a novel kind of shape-changing interaction, which was not be reported before.
Specifically, we find that the asymmetric multi-peak soliton $u_2$ exhibits shape-changing characteristics before and after interactions. Interestingly, this shape-changing
interaction involves mutual collisions of multipeak solitons as well as interactions between multipeak solitons and other-type nonlinear waves.
In particular, it is found that the shape-changing feature is specific to the multi-peak solitons.

\begin{figure}[htb]
\centering
\subfigure{\includegraphics[height=42mm,width=42mm]{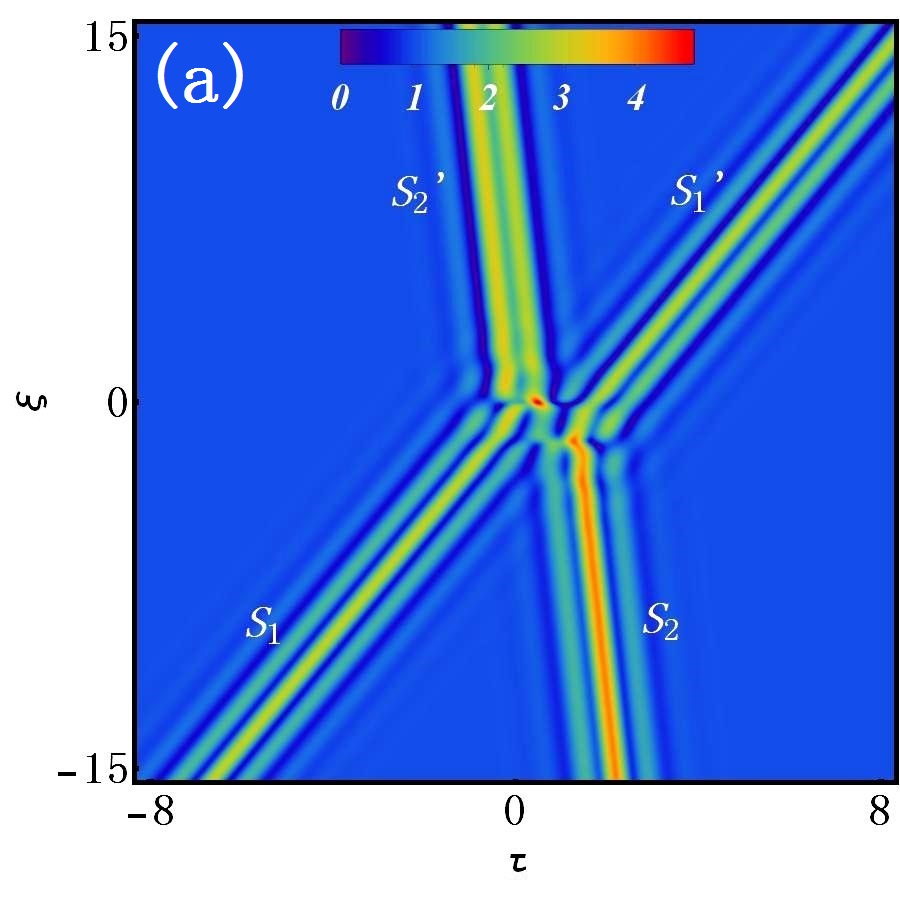}}
\hfil
\subfigure{\includegraphics[height=42mm,width=42mm]{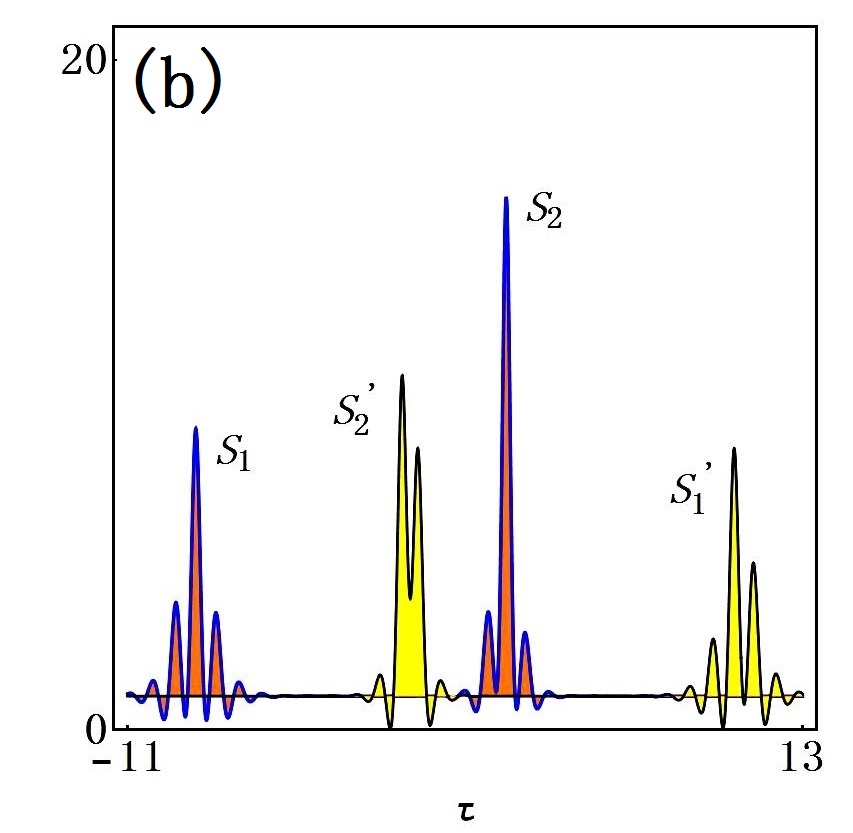}}
\caption{(color online) (a) shape-changing collision between two multi-peak solitons (incident solitons $S_1$, $S_2$; the outgoing solitons $S'_1$, $S'_2$)
with the conditions $q_j=(3q_s-q)/2$, $q\neq q_s$, $j=1,2$. (b) the corresponding intensity profiles $|u|^2$ at $\xi=-20$ (orange), $\xi=20$ (yellow).
The setup is $q=4q_s$, $a_1=1$, $a_2=1.5$, $a=1$, $\alpha=1$, $\beta=0.1$.}\label{fig6}
\end{figure}

We first study the mutual collision of two multipeak solitons with the coexistence condition $q_j=(3q_s-q)/2$, $q\neq q_s$, $j=1,2$.
As shown in Fig. \ref{fig6}, two incident multi-peak solitons $S_1$, $S_2$ with different features of peak distributions move from $\xi\rightarrow-\infty$ and
approach each other; they undergo collision around $(\xi,\tau)=(0,0)$ and form a higher peak.
They then separate with a small phase shift and propagate to $\xi\rightarrow +\infty$. It is evident to observe that the multi-peak features
of outgoing multi-peak structures $S'_1$, $S'_2$, including the peak numbers and peak intensity distributions, are changed significantly after the collision [see intensity profiles in Fig. \ref{fig6}].
This indicates that the collision is shape-changing.
For the details, by comparing intensity profiles $S_1$, $S_2$ with $S'_1$, $S'_2$,
we find that, for each soliton, the maximum intensity decreases while the sub peaks increase.
We may infer that this fascinating shape-changing collision could stem from
the intensity transfer from subpeaks to main peaks of the multi-peak soliton itself, rather than the intensity transfer between solitons.

\begin{figure}[htb]
\centering
\subfigure{\includegraphics[height=42mm,width=42mm]{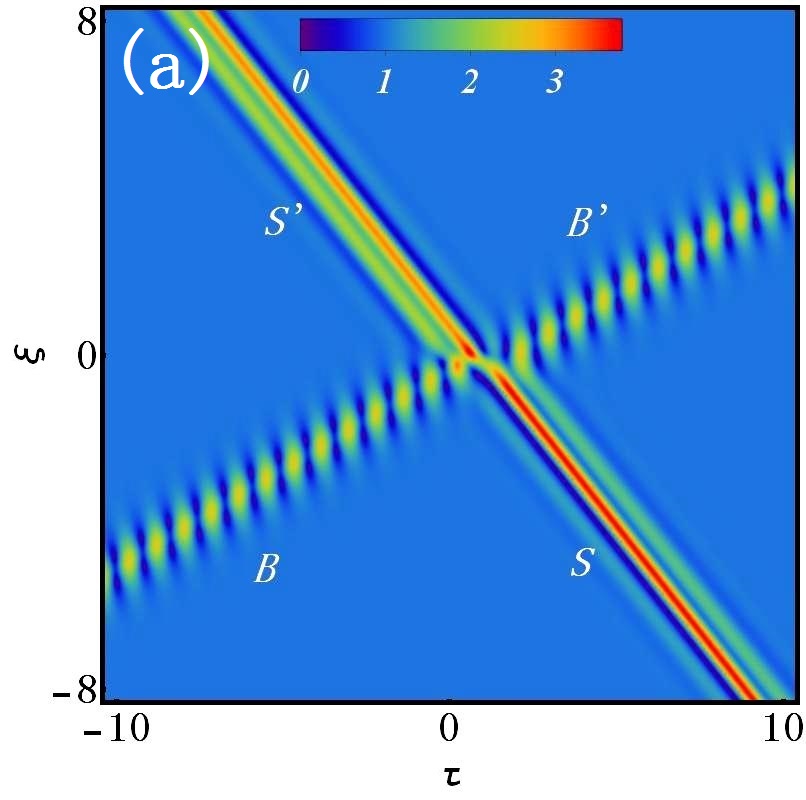}}
\hfil
\subfigure{\includegraphics[height=42mm,width=42mm]{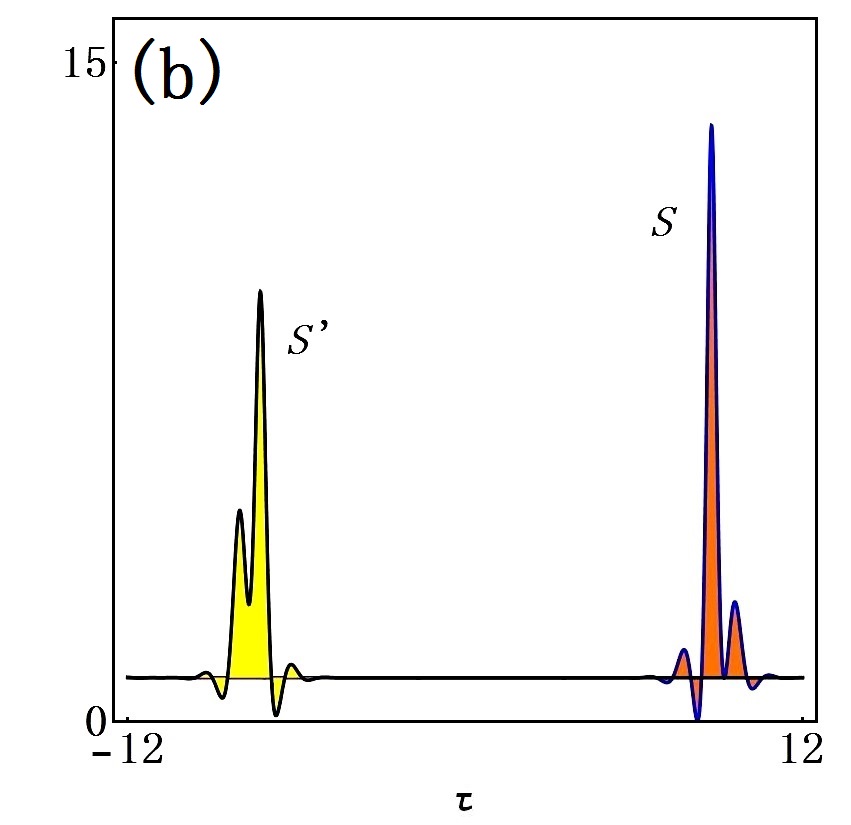}}
\caption{(color online) (a) shape-changing interaction between a breather and a multi-peak soliton (the incident soliton $S$ and breather $B$; the outgoing soliton $S'$ and breather $B'$)
with the conditions $q_1=(3q_s-q)/2$, $q\neq q_s$; $q_2\neq(3q_s-q)/2$.
(b) the corresponding intensity profiles $|u|^2$ of multi-peak solitons at $\xi=-8$ (orange), $\xi=8$ (yellow).
The setup is $q=3q_s$, $a_1=1.4$, $a_2=0.8$, $q_2=1.6$, $a=1$, $\alpha=1$, $\beta=0.1$.}\label{fig7}
\end{figure}

\begin{figure}[htb]
\centering
\subfigure{\includegraphics[height=42mm,width=42mm]{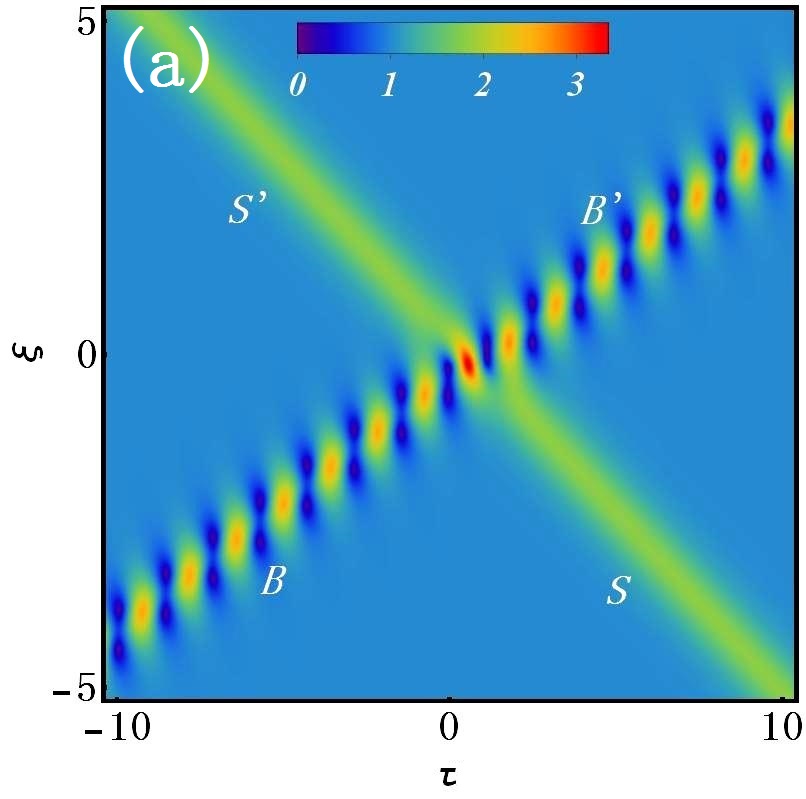}}
\hfil
\subfigure{\includegraphics[height=42mm,width=42mm]{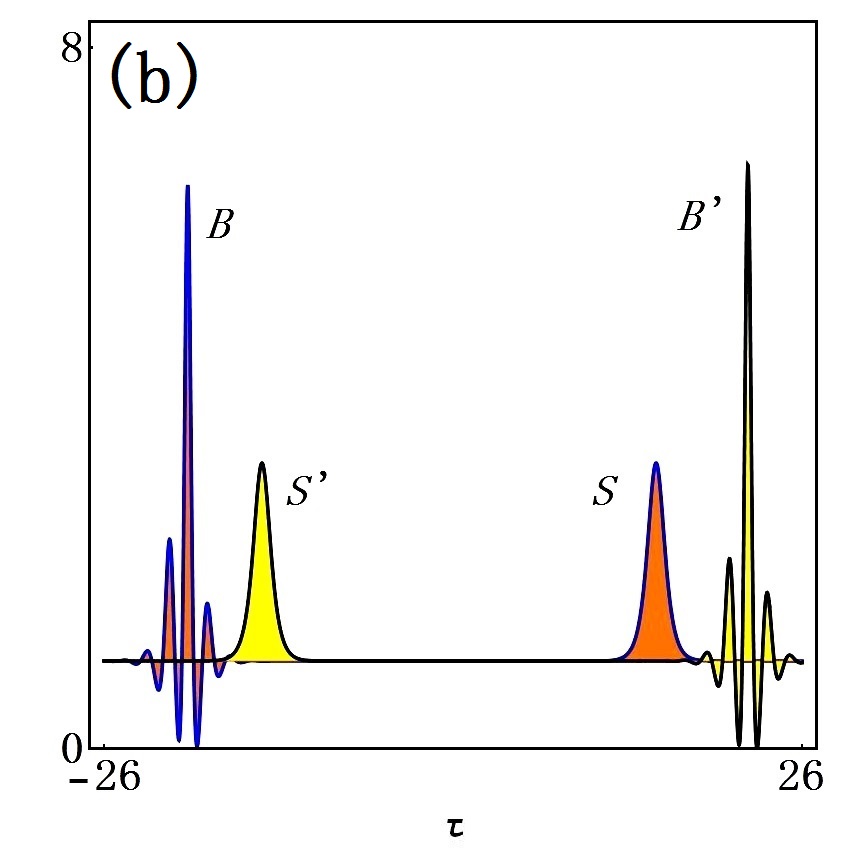}}
\caption{(color online) (a) shape-unchanging interaction between a breather and an antidark soliton (the incident soliton $S$ and breather $B$; the outgoing soliton $S'$ and breather $B'$) with the conditions $q_1=(3q_s-q)/2$, $q=q_s$, $a_1^2>a^2$; $q_2\neq(3q_s-q)/2$.
(b) the corresponding intensity profiles $|u|^2$ at $\xi=-8$ (orange), $\xi=8$ (yellow).
The setup is the same as in Fig. \ref{fig7}, but $q=q_s$.}\label{fig8}
\end{figure}

To prove the validity of the conjecture above, we then analyze the localized energy of light pulse against the background of multipeak solitons with the expression $I_e=\int_{\tau_1}^{\tau_2}\left\{|u(\xi,\tau)|^2-a^2\right\}d\tau$.
As a result, the localized energy of multi-peak solitons before and after the shape-changing collision can be investigated via optimal numerical method at different propagation distance $\xi$.
The total localized energy of these two solitons [$I_e(s_1)+I_e(s_2)$] is conserved which can be verified directly when $\tau_1=-\infty$, $\tau_2=+\infty$.
The localized energy of the each multi-peak soliton [$I_e(s_1)$, $I_e(s_2)$] is obtained by the appropriate choice for $\tau_1$, $\tau_2$.
For a selected initial propagation distance, i.e., $\xi<0$, we calculate $I_e(s_1)$ and $I_e(s_2)$ by $\int_{-\infty}^{\tau_0}\left\{|u|^2-a^2\right\}d\tau$ and  $\int_{\tau_0}^{+\infty}\left\{|u|^2-a^2\right\}d\tau$, respectively. Note that the transverse position $\tau_0$ is located between $S_1$, $S_2$, leading to $|u(\xi,\tau_0)|= a$, and vice versa for positive $\xi$.
Then the interesting finding is that each multi-peak soliton preserves its localized energy before and after the collision.
Namely, after multi-peak solitons collide, the shape change and localized energy conservation of the each multi-peak soliton coexist.
It is noted that, this shape-changing feature is completely different from the known shape-changing collisions between standard solitons in the coupled
NLS systems that describe a process of energy transference between solitons in one component \cite{t}.

For a better understanding of the characteristic of this shape-changing characteristic of multi-peak solitons, we
next explore interspecific interactions, i.e., the interactions between multi-peak solitons and other-type localized nonlinear waves (breathers).
Our aim is to demonstrate that after the interspecific interaction occurs, the multi-peak solitons exhibit intensity redistribution
but the characteristic of breathers remains invariant.

Figure \ref{fig7} illustrates the interaction between multi-peak solitons and breathers for the choice of the parameters
$q_1=(3q_s-q)/2$, $q\neq q_s$, $q_2\neq(3q_s-q)/2$. One can see from Fig. \ref{fig7} that, an incident multi-peak soliton $S$ propagating along $\xi$ collides with a breather
near $\xi=0$; after that the outgoing soliton $S'$ shows a typical shape-changing characteristic but the breather remains the original feature.
An analysis of the intensity profiles and the localized energy of $S$ and $S'$ shows that, after the interspecific interaction with a breather occurs, the single multi-peak soliton allows the intensity redistribution between their peaks, and preserves strictly the localized energy of the soliton.

Finally, we illuminate that the unique shape-changing characteristic is specific to the multi-peak solitons.
To this end, we consider the limiting case of the interaction shown in Fig. \ref{fig7}, i.e., the collision between antidark solitons and breathers.
The corresponding interaction structure is displayed in Fig. \ref{fig8} with the limiting condition $q=q_s$.
It is shown evidently that this collision exhibits a completely shape-unchanging feature although the waves are two distinct types of localized waves.
As a result, a comparison of Figs. \ref{fig7} and \ref{fig8} indicates that, the shape-changing feature is specific to asymmetric multi-peak solitons and is not available for antidark solitons.

\section{Conclusion}
In summary, symmetric and asymmetric multi-peak solitons on a continuous wave background in the femtosecond regime have been investigated analytically and numerically.
Key properties of such multi-peak solitons, as the formation mechanism, propagation stability, and shape-changing collisions, have been revealed in detail, for the first time to our knowledge.

This intriguing multi-peak soliton exhibits both localization and periodicity along the transverse distribution on a background. The corresponding periodicity and localization for multi-peak (symmetric or asymmetric) solitons are well described by a periodic wave and a single-peak (W-shaped or antidark) soliton, respectively.

Although the maximum optical intensity is different,
the interesting connection is that, the optical intensity against the background of the symmetric and asymmetric solitons, turns out to coincide with each other under the same initial parameter condition, i.e.,
$\int_{-\infty}^{+\infty}\left(|u_{1}|^2-a^2\right)\textrm{d}\tau=\int_{-\infty}^{+\infty}\left(|u_{2}|^2-a^2\right)\textrm{d}\tau$.

Especially, a phase diagram for different types of nonlinear excitations on a continuous wave background including breather, rogue wave, W-shaped soliton, antidark soliton, periodic wave, and multi-peak soliton is established based on the explicit link between exact nonlinear wave solution and MI analysis.
Numerical simulations were performed to confirm the propagation stability of the multi-peak solitons with symmetric and asymmetric structures.

Finally, the remarkable shape-changing feature of asymmetric multi-peak solitons, occurring not only in the intraspecific collision (soliton mutual collision) but also in the interspecific interaction (soliton-breather interaction), was unveiled. It is demonstrated that each multi-peak soliton exhibits the coexistence of shape change and conservation of the localized energy of light pulse against the continuous wave background.
The shape-changing interaction between nonlinear waves on a continuous wave background will enrich our understanding of localized wave collision in the (1+1)-dimensional scalar nonlinear wave evolution systems.

\section*{ACKNOWLEDGEMENTS}
This work has been supported by the National Natural Science Foundation of China (NSFC)(Grant Nos. 11475135, 11547302, 61505101),
and the ministry of education doctoral program funds (Grant No. 20126101110004). We are grateful to Prof. Q. Guo (South China Normal University) for his valuable comments at NSOS (Nov.06-08, 2015). C. Liu also thanks Prof. F. Baronio for his helpful suggestion and Fa-kai Wen, Wen-Hao Xu for their helpful discussions.

\appendix
\renewcommand\appendixname{APPENDIX}
\section{}
A general nonlinear wave solution on the background $u_0$ of the higher-order NLSE (\ref{equ1}), which describes optical femtosecond pulse propagation in a single-mode fiber, is present as follows
\begin{equation}
\label{equ:g}
u_{1,2}=\left[\frac{\Delta_{1,2}\cosh(\varphi+\delta_{1,2})+\Xi_{1,2}\cos(\phi+\xi_{1,2})}
{\Omega_{1,2}\cosh(\varphi+\omega_{1,2})+\Gamma_{1,2}\cos(\phi+\gamma_{1,2})}+a\right]e^{i\theta},
\end{equation}
where
\begin{eqnarray}
\varphi&=&2\eta_i (\tau + V_1\xi),~~~\phi = 2\eta_r(\tau + V_2\xi),\nonumber\\
V_1&=&v_1 + v_2\eta_r/\eta_i,~~~V_2=v_1-v_2\eta_i/\eta_r,\nonumber\\
v_1&=&\beta(2 a^2+4 a_1^2- q'^2)-(q_1+q)(q\beta+\alpha/2),\nonumber\\
v_2&=&a_1[\alpha + 2\beta(q+2 q_1)],~~~\eta_r+i\eta_i=\sqrt{\epsilon+i\epsilon'},\nonumber\\
\epsilon&=&a^2-a_1^2 + (q-q_1)^2/4,~~~\epsilon'=a_1(q-q_1),\nonumber
\end{eqnarray}
Here we remark that if $V_1=V_2$, implying $v_2=0$ [thus $q_1=-\alpha/(4\beta)-q/2$], the solution \ref{equ:g} describes the dynamics of multi-peak solitons in Eq. (\ref{equ2}).
Instead, if $V_1\neq V_2$, implying $v_2\neq0$, the solution \ref{equ:g} displays the properties of the breather and rogue wave.
Specifically, with the condition $0<a_1<a$, $q=q_1$, the Akhmediev breather solution is given with the form
\begin{equation}
u_{1,2}=\left[\frac{2\eta^2\cosh(\kappa \xi)+i 2\eta a_1 \sinh(\kappa \xi)}{a\cosh(\kappa \xi)-e^{i\sigma}a_1 \cos[2\eta(\tau+v_1\xi)-\mu]}-a\right]e^{i\theta},
\end{equation}
where $v_1=\beta(2 a^2+4 a_1^2- q^2)-2q(q\beta+\alpha/2)$, $\kappa=2\eta v_2$, $v_2=a_1\alpha(1-q/q_s)$, $q_s=-\alpha/(6\beta)$, $\sigma=\sigma_{1,2}=\{0,\pi\}$ with $\mu=\mu_{1,2}=\{0,\arctan(-\eta_r/a_1)\}$. The AB is periodic along the distribution direction $\tau$, and the period is $D_\tau=\pi/\sqrt{a^2-a_1^2}$. On the other hand, the Kuznetsov-Ma breather is obtained with the condition $a_1>a$, $q=q_1$,
\begin{equation}
u_{1,2}=\left[\frac{2\eta'^2\cos(\kappa' \xi)+i2\eta' a_1\sin(\kappa' \xi)}{e^{i\sigma}a_1 \cosh[2\eta'(\tau+v_1\xi)+\mu']-a\cos(\kappa' \xi)}-a\right]e^{i\theta},
\end{equation}
where $\kappa'=2\eta' v_2$, $\sigma=\sigma_{1,2}=\{0,\pi\}$ with $\mu'=\mu'_{1,2}=\{0,\textrm{arctanh}(-\eta_i/a_1)\}$. Note that if $q=q_s$, the AB and KMB are converted to the periodic wave and W-shaped/antidark soliton described by Eqs. (\ref{equ4}) and (\ref{equ5}). For clarity, various types of nonlinear excitations extracting from the general solution on a background is classified in the following table,
\begin{table}[!hbp]
 \label{table1}
 \begin{tabular}{|c|c|}
 \hline
  Nonlinear waves type  & Existence condition \\
  \hline
  Breather and rogue wave & $q_j\neq\frac{3q_s-q}{2}$~($q_s=\frac{-\alpha}{6\beta}$, $j=1,2$)\\
  \hline
  Multi-peak soliton & $q_j=\frac{3q_s-q}{2}$, $q\neq q_s$ \\
  \hline
  W-shaped/Antidark soliton & $q_j=\frac{3q_s-q}{2}$, $q=q_s$, $a^2<a_j^2$ \\
  \hline
  Periodic wave &  $q_j=\frac{3q_s-q}{2}$, $q=q_s$, $a^2>a_j^2$ \\
  \hline
  Rational W-shaped soliton &  $q_j=\frac{3q_s-q}{2}$, $q=q_s$, $a^2=a_j^2$ \\
  \hline
\end{tabular}
\caption{Types of nonlinear excitations and corresponding explicit existence conditions.}
\end{table}

\end{document}